\documentstyle[twocolumn,prl,aps,epsf]{revtex}
\begin{document}
\title{Determining the sign of $\Delta_{31}$ at long baseline neutrino
experiments}
\author
{Mohan Narayan and S. Uma Sankar}  
\address
{Department of Physics, I.I.T. , Powai, Mumbai 400076, India}
\date{\today}
\maketitle

\begin{abstract} 
Recently it is advocated that high intensity and low energy 
$(E_\nu \sim 2~{\rm GeV})$ neutrino beams should be built to 
probe the $(13)$ mixing angle $\phi$ to a level of a few parts
in $10^4$. Experiments using such beams will have better signal
to background ratio in searches for $\nu_\mu \rightarrow \nu_e$ 
oscillations. We propose that such experiments can also 
determine the sign of $\Delta_{31}$ even if the beam consists
of {\it neutrinos} only. By measuring the $\nu_\mu \rightarrow
\nu_e$ transitions in two different energy ranges, the 
effects due to propagation of neutrinos through earth's crust 
can be isolated and the sign of $\Delta_{31}$ can be determined.
If the sensitivity of an experiment to $\phi$ is $\varepsilon$, 
then the same experiment is automatically sensitive to matter 
effects and the sign of $\Delta_{31}$ for values of $\phi 
\geq 2 \varepsilon$. 
\end{abstract} 
\vspace{0.5cm}

Neutrino oscillations provide an elegant explanation for the 
observed deficits in solar and atmospheric neutrino fluxes. 
Both these deficits can be accounted for
in a three flavor oscillation scheme. In this scheme, the oscillation
probabilities, in general, depend on six paramaters: two independent 
mass-squared differences $\Delta_{21}$ and $\Delta_{31}$, three mixing
angles $\omega, \phi$ and $\psi$ and a CP violating phase. Given the 
distance and energy scale of solar neutrinos, their oscillations depend
only on one mass-squared difference $\Delta_{21}\sim 10^{-5}$ eV$^2$ 
and two mixing angles $\omega$ and $\phi$ \cite{kupa,ajmm}. 
Atmospheric neutrino oscillations are insensitive to a mass-squared 
difference as small as $10^{-5}$ eV$^2$. Hence $\Delta_{21}$ can be
set to zero in the analysis of atmospheric neutrinos. In this 
approximation, the oscillation probabilities of atmospheric neutrinos
depend on $|\Delta_{31}| \sim (1-7) \times  10^{-3}$ eV$^2$ \cite{superk}
and two mixing angles $\phi$ and $\psi$ \cite{jim94}. The results of 
CHOOZ experiment \cite{chooz} give the strong constraint 
$\sin^2 2 \phi \leq 0.1$. 
This implies that either $\phi \leq 9^\circ$ or $\phi \geq 81^\circ$.
The second possibility is ruled out by the constraints from both 
solar neutrino \cite{mnsu} and atmospheric neutrino analyses 
\cite{lipari}. Thus $\phi$ is small and the atmospheric
neutrino oscillations are almost $\nu_\mu \rightarrow \nu_\tau$             
oscillations with a small $\nu_\mu \leftrightarrow \nu_e$ component. 
To explain the observed deficit of muon neutrinos, the mixing angle
$\psi$ should be close to $\pi/4$ \cite{superk}.

Combining the data from solar, atmospheric neutrino
experiments and CHOOZ we have the following pattern of neutrino masses 
and mixings: Two mass eigenstates are very close together and the third
mass eigenstate is a little apart. This third state contains
approximately equal admixture of $\nu_\mu$ and $\nu_\tau$ and a small 
admixture of $\nu_e$. From a model building point of view, the crucial 
question is whether this third state is more massive than the two nearly 
degenerate states ($\Delta_{31}$ positive) or less massive ($\Delta_{31}$
negative).
 
Long baseline experiments are designed to observe $\nu_\mu$ oscillations
under controlled conditions. The baselines and energies of these
experiments are chosen such that significant deficit of $\nu_\mu$ flux
will be observed if the neutrino parameters are in the range   
determined by Super-Kamiokande atmospheric neutrino analysis.
Two different long baseline neutrino beams, one from Fermilab to
Soudan \cite{mintdr} and another from CERN to Gran Sasso 
\cite{arubbia}, are being constructed. In both cases the baseline
length is 730 km and the beam consists of neutrinos only. The experiments
downstream expect to observe neutrino oscillations via i) $\nu_\mu$ 
disappearance ii) $\nu_\tau$ appearance and iii) $\nu_e$ appearance.
The high stastistics of these experiments allow them to determine
the energy distribution of $\nu_\mu$ charged current (CC) events. 
By locating the
minimum of $\nu_\mu$ CC event distribution, one can determine 
$|\Delta_{31}|$ and $\sin^2 2 \psi$ independently and to a precision of
$10 \%$ \cite{petyt}. But neither $\nu_\mu$ disappearance nor $\nu_\tau$
appearance data can give us information on the sign of $\Delta_{31}$. 

$\nu_\mu \rightarrow \nu_e$ oscillation probability in long baseline 
experiments is modified by the propagation of neutrinos through the
matter of earth's crust. Matter effects \cite{wolf} boost the 
oscillation probability for neutrinos if $\Delta_{31}$ is positive
and suppress it if $\Delta_{31}$ is negative. The situation is 
reversed for anti-neutrinos. At neutrino factories, where $\nu_\mu$
and $\bar{\nu}_\mu$ beams from muon storage rings will be available, 
one can measure $\nu_\mu \rightarrow \nu_e$ and $\bar{\nu}_\mu
\rightarrow \bar{\nu}_e$ oscillation probabilities, 
isolate the difference induced by the matter effects and determine
the sign of $\Delta_{31}$ \cite{bgrw}. However, the neutrino beam
energies at these factories will be very high ($E_\nu \sim 20$ GeV)
and hence one needs a very long baseline (of about $7000$ km) so that
the oscillation probability will not be too small. It may be a long  
time before this ambitious program is realized. Recently  
it was proposed that high intensity, low energy neutrino beams can 
be built with conventional techniques \cite{richter}. If these beams
have peak flux around $E_\nu \sim 2$ GeV, then a 20 Kiloton detector 
at a distance of about $730$ km will be capable of measuring values of  
$\phi$ as small as a few parts in $10^{4}$. In this letter we show that 
such a detector can determine the sign of $\Delta_{31}$ even if
the beam consists of {\it neutrinos only}. 

In three flavor scheme, $\nu_\mu \rightarrow \nu_e$ oscillation 
probability is given by
\begin{equation}
P_{\mu e} = \sin^2 \psi \sin^2 2 \phi \sin^2 \left( \frac{1.27  \
\Delta_{31} L}{E} \right), \label{eq:vacP}
\end{equation}
where $\Delta_{31}$ is in eV$^2$, the baseline length $L$ is in km and 
the neutrino energy $E$ is in GeV. From CHOOZ constraint, we see that 
this probability is atmost a few percent. In the long baseline  
experiments, the neutrinos propagate through earth's crust which
has constant density of about $3$ gm/cc. The oscillation probability 
with the inclusion of matter effects is given by \cite{mnsu,lipari}
\begin{equation}
P^m_{\mu e} = \sin^2 \psi \sin^2 2 \phi^m \sin^2 \left( \frac{1.27  \
\Delta^m_{31} L}{E} \right). \label{eq:matP}
\end{equation}
Comparing Eq.~(\ref{eq:vacP}) with Eq.~(\ref{eq:matP}), we note that
the mixing angle $\psi$ is unaffected by matter term and $\phi$ and 
$\Delta_{31}$ are replaced by their matter modified values $\phi^m$
and $\Delta^m_{31}$ respectively \cite{jim94,mnsu}. These are given by
\begin{eqnarray}
\sin 2 \phi^m & = & \frac{\Delta_{31} \sin 2 \phi}{\Delta^m_{31}},    
\nonumber \\
\Delta^m_{31} & = & \sqrt{(\Delta_{31} \cos 2 \phi - A)^2 +
(\Delta_{31} \sin 2 \phi)^2}.
\end{eqnarray}
Here $A$ is the matter term and is given by
\begin{equation}
A = 2 \sqrt{2} G_F N_e E 
= 0.76 \times 10^{-4} ~~\rho~({\rm in~gm/cc})~E~({\rm in~GeV}).
\end{equation}
If $\Delta_{31}$ is positive, a resonance will occur for neutrinos 
at the energy
\begin{equation}
E_{res} = \frac{\Delta_{31} \cos 2 \phi}{2.3 \times 10^{-4}}.
\end{equation}
At this energy, $\Delta_{31} \cos 2 \phi = A$
and $\sin 2 \phi^m = 1$ \cite{mism}. Unfortunately, for a baseline
of $730$ km, $P^m_{\mu e}$
at $E_{res}$ is not significantly greater than $P_{\mu e}$ at the 
same energy. This occurs because, at resonance, $\sin 2 \phi^m$ is  
maximized but simultaneously $\Delta^m_{31}$ is minimized \cite{ajmm}.  
Hence nothing is gained by tuning the neutrino flux to peak around 
$E_{res}$. Away from the resonance, the energy dependence of $\sin^2 
2 \phi^m$ is different from that of $\sin^2 (1.27 \Delta^m_{31}L/E)$. 
So, in general, we can expect $P^m_{\mu e}$ to be different from 
$P_{\mu e}$. In Figure~1, we plotted $P_{\mu e}$ (middle line), 
$P^m_{\mu e}$ with $\Delta_{31}$ positive (upper line) and  
$P^m_{\mu e}$ with $\Delta_{31}$ negative (lower line). We chose  
$L= 730$ km, $\phi = 9^\circ$ and $|\Delta_{31}| = 3.5 \times 10^{-3}$  
eV$^2$, which is Super-Kamiokande best-fit value.
We see that for both signs of 
$\Delta_{31}$, the maximum of $P^m_{\mu e}$ occurs close to the energy 
where $P_{\mu e}$ is maximum, that is where the phase of the oscillating  
term $1.27 |\Delta_{31}| L/E = \pi/2$. We call this energy to be 
\begin{equation}
E_{\pi/2} = \frac{2.54~|\Delta_{31}| L}{\pi}.
\end{equation}
For $|\Delta_{31}| = 3.5 \times 10^{-3}$ eV$^2$, $E_{\pi/2} = 2$ GeV,
if $L = 730$ km.

Long baseline experiments will first measure $|\Delta_{31}|$ and 
$\sin^2 2 \psi$ to about $10 \%$ precision. The next goal of 
neutrino experiments is to probe small values of $\phi$ by 
looking for $\nu_\mu \rightarrow \nu_e$ oscillations. In such
a search, the neutrino beam energy should be tuned such that 
the oscillation signal to background ratio is maximized. 
Neutrino crosssections increase linearly with neutrino energy and
neutrino beam flux increases with the energy of the intial 
accelerated particle so neutrino event rates increase as $E^p$
where $p > 1$. So, it seems as if high energy beams are 
more favorable to do neutrino physics than low energy beams.
However, $\nu_\mu \rightarrow \nu_e$ oscillation probability
falls as off as $E^{-2}$ at energies much greater than $E_{\pi/2}$.  
Hence the increase in signal event rate, if any, is moderate.      
The background events come primarily from the neutral current 
events of neutrinos in the beam and the CC events
of the $1 \%$ $\nu_e$ contamination of the beam. Both these 
are proportional to total event rate and increase as $E^p$. 
Hence the $\nu_\mu \rightarrow \nu_e$ signal to background 
ratio worsens as $E^{-2}$ if one goes to energies $E >>  
E_{\pi/2}$. In order to maximize this signal, it is 
advantageous to tune the energy such that flux of the neutrino 
beam peaks around $E_{\pi/2}$. A recent paper by B. Richter 
\cite{richter} also proposes such a tuning, where it is also 
advocated that construction of high intensity and low energy 
neutrino beams can be the next step in the effort to measure small 
neutrino parameters. At such beams the signal to background ratio
for $\nu_\mu \rightarrow \nu_e$ oscillations, will be much larger 
than in the case of high energy beams. Such beams can be used to 
probe values of $\phi$ as small as a few parts in $10^4$. They  
can also be used to determine the sign of $\Delta_{31}$, even if
they consist of neutrinos only.

We note from Figure~1 that $P^m_{\mu e}$ peaks
close to where $P_{\mu e}$ peaks, that is close to $E = E_{\pi/2}$.
In addition, we can make three important observations:
\begin{enumerate}
\item
Above $4$ GeV (or above $2 E_{\pi/2}$), matter effects have negligible 
effect on $\nu_\mu \rightarrow \nu_e$ oscillation probability. For both
signs of $\Delta_{31}$, $P^m_{\mu e}$ is very close to $P_{\mu e}$
for energies above $2 E_{\pi/2}$. Hence by selecting events with
$E > 2 E_{\pi/2}$, {\it one can directly measure $\phi$
without worrying about matter effects}.
\item
Matter effects are most dominant in the neighbourhood of $E_{\pi/2}$.
In this neighbourhood, $P^m_{\mu e}$ is about $25 \%$ greater than 
$P_{\mu e}$ if $\Delta_{31}$ is positive and is smaller by about the 
same factor if $\Delta_{31}$ is negative. This is true for all values 
of $|\Delta_{31}|$ and $\phi$ \cite{lipari}. 
\item
For fixed value $|\Delta_{31}|/E$, matter effects cause a larger change 
for larger
values of $L$. This occurs for the following reason. In the 
neighbourhood of $E_{\pi/2}$, the phase of the oscillating term
is $\pi/2$. However, $E_{\pi/2}$ is larger if $L$ is larger and
$\sin^2 2 \phi_m$ is larger for larger energies. So matter term
causes a larger change in $P^m_{\mu e}$ around $E_{\pi/2}$ for larger 
values of $L$.
\end{enumerate}
To observe the differences induced by matter effects, one needs to 
measure $\nu_\mu \rightarrow \nu_e$ oscillation probability in the
lower energy range $0 < E_\nu < 2 E_{\pi/2}$. Using the value of 
$\phi$ measured from the data of the higher energy range 
$E_\nu > 2 E_{\pi/2}$, one can predict the expected number
of events for the lower energy range,  
for $\Delta_{31}$ positive and 
for $\Delta_{31}$ negative. By comparing the results of the lower energy
range with the predictions, one can determine the sign of $\Delta_{31}$.
Since the vacuum value of $\phi$ is unknown, one needs a minimum of
two measurements to determine the sign of $\Delta_{31}$. Usually 
these are taken to be $\nu_\mu \rightarrow \nu_e$ and $\bar{\nu}_\mu
\rightarrow \bar{\nu}_e$ oscillation probabilities. But, as we 
argued above, $\nu_\mu \rightarrow \nu_e$ event samples in the lower 
and higher energy ranges 
also can be used to distinguish between the two possible signs of 
$\Delta_{31}$. The neutrino beam should have reasonable width in energy 
(say $0 < E_\nu < 4 E_{\pi/2}$)
so that there will be substantial number of events both in the lower
and higher energy regions. 

The following procedure may be used to determine the sign of 
$\Delta_{31}$. From the experiment we will get two pieces of data, 
$N_{eh}$ from the higher energy region and $N_{el}$ from the lower 
energy region. We wish to check whether the 
hypothesis of positive or negative $\Delta_{31}$ fits the data  
better. Let $N_{eh}^{m+} (\phi)$ and $N_{eh}^{m-} (\phi)$ are the 
theoretical expectation for the number of electron events in the
higher energy region for positive and negative $\Delta_{31}$ 
respectively.
These two numbers will be functions of the mixing angle $\phi$ which,
as yet, is unknown. 
Similarly, $N_{el}^{m+} (\phi)$ and $N_{el}^{m-} (\phi)$ are the 
theoretical expectations for the lower energy region.
Now we define two different $\chi^2$s,
\begin{eqnarray}
\chi_+^2 & = & \frac{(N_{eh} - N_{eh}^{m+})^2}{N_{eh}} +
 \frac{(N_{el} - N_{el}^{m+})^2}{N_{el}}, \nonumber \\ 
\chi_-^2 & = & \frac{(N_{eh} - N_{eh}^{m-})^2}{N_{eh}} +
 \frac{(N_{el} - N_{el}^{m-})^2}{N_{el}}. 
\end{eqnarray}
Both these $\chi^2$s are functions of $\phi$ and by varying $\phi$ 
they can be minimized. If $\Delta_{31}$ is positive, we will have
$\chi^2_+|_{min} << \chi^2_-|_{min}$ and vice-verse if $\Delta_{31}$ 
is negative. This strong inequality between the respective minima of
the two $\chi^2$s must occur if our present understanding of effect
of matter term on the propagation of neutrinos is correct.

Let us briefly consider the ability of long baseline neutrino 
experiments to determine the sign of $\Delta_{31}$. Suppose 
the minimum value of $\phi$ an experiment can measure is 
$\varepsilon$. 
In the following, we assume that the neutrino beam has the 
same spectrum as that of the low energy option of MINOS \cite{petyt}.
We will also take $|\Delta_{31}| = 3.5 \times 10^{-3}$ eV$^2$ 
and the baseline length to be $L = 730$ km, so that $E_{\pi/2} =2$ 
GeV. In such a case, the 
$\nu_\mu \rightarrow \nu_e$ oscillation signal events are 
split $3:1$ between the lower energy range ($0-4$ GeV)
and the high energy range ($> 4$ GeV). The $\nu_\mu$ flux,
and hence the background events, are split in the ratio
$4:6$ for the same two ranges. The signal to background ratio
in the lower energy range is about 5 times better than in the
higher energy range. This illustrates our earlier point
that neutrino beams with energy tuned to $E_{\pi/2}$ are better 
than high energy beams
for observing $\nu_\mu \rightarrow \nu_e$ oscillations. 
Let us assume that, to be sensitive  
to $\phi = \varepsilon$, the experiment must be capable
of observing $N$ $\nu_e$-CC events above the background.
Of these, $3N/4$ will be in the lower energy range and $N/4$
will be in the higher energy range. If the value of $\phi$ is
$2 \varepsilon$, then this experiment will see $4N$ 
$\nu_e$-CC events in case of vacuum $\nu_\mu \rightarrow \nu_e$
oscillations. Of these, $3N$ will be in the lower energy range
and $N$ will be in the higher energy range. The number of events
in the higher energy range is not affected by matter effects and 
this number, $N$, is large enough to be detectable above the 
background. Matter effects boost the events in the lower energy 
range to $3.75 N$ if $\Delta_{31}$ is positive and suppress them 
to $2.25 N$ if $\Delta_{31}$ is negative. In each case, the 
difference induced by the matter effects is larger than the 
background in this energy range, which is less than half the total
background. Hence, any long baseline neutrino experiment which
sensitive to $\phi \simeq \varepsilon$ is automatically sensitive
to the sign of $\Delta_{31}$ if $\phi \simeq 2 \varepsilon$.

In conclusion, we showed that neutrino beams with energy tuned to
$E_{\pi/2}$ have better
signal to background ratios than the high energy neutrino beams.
We also showed that a future high intensity and low energy neutrino
experiment, which can measure values of $(13)$ mixing angle $\phi$ as
small as a few parts in $10^4$, can automatically measure the sign of 
$\Delta_{31}$ for values of $\phi$ greater than double the sensitivity
of the experiment even if the beam consists of neutrinos only. 

We would like to thank Prof. A. Bettini and K.R.S. Balaji for discussions. 
We would also like to thank the Department of Science and Technology, 
Govt. of India for supporting this research.

\begin{figure}[ht]
\begin{center}
\leavevmode\epsfysize=6cm \epsfbox{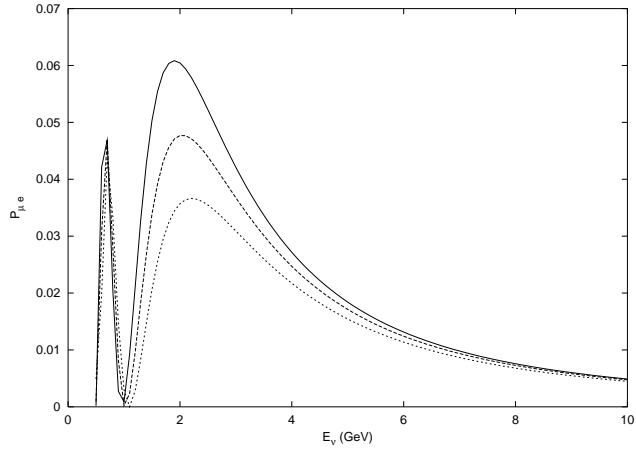}
\end{center}
\caption{$\nu_\mu \rightarrow \nu_e$ oscillation probabilities vs $E$ 
for $|\Delta_{31}| = 3.5 \times 10^{-3}$ eV$^2$, $\sin^2 2 \phi = 0.1$
and $L = 730$ km.
The middle line is $P_{\mu e}$, the upper line is $P^m_{\mu e}$ with  
$\Delta_{31}$ positive and the lower line is $P^m_{\mu e}$ with 
$\Delta_{31}$ negative.}
\end{figure}

\begin{references}
\bibitem{kupa}
T. K. Kuo and J. Pantaleone, Phys. Rev. D{\bf 35}, 3432 (1987).
\bibitem{ajmm}
A. S. Joshipura and M. V. N. Murthy, Phys. Rev. D{\bf 37}, 1374 (1988).
\bibitem{superk}
H. Sobel, Talk presented at {\it Neutrino 2000}, Sudbury, Canada, 
June 2000.
\bibitem{jim94}
J. Pantaleone, Phys. Rev. D{\bf 49}, 2152 (1994).
\bibitem{chooz}
CHOOZ Collaboration: M. Appollonio {\it et al}, Phys. Lett. {\bf 466B},
415 (1999).
\bibitem{mnsu}
Mohan Narayan and S. Uma Sankar, Phys. Rev. D{\bf 61}, 013003 (2000).
\bibitem{lipari}
P. Lipari, Phys. Rev. D{\bf 61}, 113004 (2000).
\bibitem{mintdr}
MINOS technical Design Report, Nu-MI-L-337, October 1998.
\bibitem{arubbia}
A. Rubbia, Talk presented at {\it Neutrino 2000}, Sudbury, Canada, 
June 2000.
\bibitem{petyt}
David A. Petyt, {\it Physics Potential of the three Ph2 beam 
designs}, MINOS preprint Nu-MI-L 612 (April, 2000).
\bibitem{wolf}
L. Wolfenstein, Phys. Rev. D{\bf 17}, 2369 (1978); {\it ibid}
D{\bf 20}, 2634 (1979).
\bibitem{bgrw}
C. Albright {\it et al}, {\it Physics at Neutrino Factory}, 
hep-ex/0008064
\bibitem{richter}
B. Richter, SLAC-PUB-8587, hep-ph/0008222 
\bibitem{mism}
S. P. Mikheyev and A. Yu. Smirnov, Yad. Fiz. {\bf 42}, 1441 (1985)
[Sov. J. Nucl. Phys. {\bf 42}, 913 (1985)]; Nuovo Cimento {\bf C9},
17 (1986).
\end{references}
\end{document}